\documentclass[11pt]{article} 

\usepackage[utf8]{inputenc} 
\usepackage[T1]{fontenc}
\usepackage{currvita}
\usepackage[noblocks]{authblk}     

\usepackage{graphicx}
\usepackage{caption}
\usepackage{subcaption}

\usepackage{geometry} 
\usepackage{booktabs} 
\usepackage{array} 
\usepackage{paralist} 
\usepackage{verbatim} 
\usepackage{subfig} 
\usepackage{url}

\usepackage{fancyhdr} 
\usepackage{sectsty}
\allsectionsfont{\sffamily\mdseries\upshape} 

\usepackage[nottoc,notlof,notlot]{tocbibind} 
\usepackage[titles,subfigure]{tocloft} 

\title{Conspiratorial beliefs observed through entropy principles}
\author{Nata\v{s}a Golo and Serge Galam}
\affil{Cevipof, Center for political research, SciencesPo and CNRS, Paris, France} 

\begin{document}
\maketitle

{\bf Abstract}: We propose a novel approach framed in terms of information theory and entropy to tackle the issue of conspiracy theories propagation. We start with the report of an event (such as 9/11 terroristic attack) represented as a series of individual strings of information denoted respectively by two-state variable $E_{i}=\pm1, i=1,…, N$.  Assigning $E_{i}$ value to all strings, the initial order parameter and entropy are determined. Conspiracy theorists comment on the report, focusing repeatedly on several strings $E_{k}$ and changing their meaning (from $-1$ to $+1$). The reading of the event is turned fuzzy with an increased entropy value. Beyond some threshold value of entropy, chosen by simplicity to its maximum value, meaning $N/2$ variables with $E_{i}=1$, doubt prevails in the reading of the event and the chance is created that an alternative theory might prevail. Therefore, the evolution of the associated entropy is a way to measure the degree of penetration of a conspiracy theory. Our general framework relies on online content made voluntarily available by crowds of people, in response to some news or blog articles published by official news agencies. We apply different aggregation levels (comment, person, discussion thread) and discuss the associated patterns of entropy change.

\section{Introduction}
\label{sec:1}

While conspiracy theories have existed for centuries, the phenomenon stayed confined to small groups of people due to the slow process of spreading based mainly on word-of-mouth propagation. However, the recent use of internet by conspiracists has altered the nature of the phenomenon, providing both the increased speed of propagation and the dramatic extension of people exposed to the rumor. Moreover, in the new context of global terrorism, the existence of conspiracy theories disseminated in the population became an issue in the ability of democratic societies to fight against terrorism. 
The propagation of conspiratorial beliefs has been previously renowned in \cite{bib1}, \cite{bib2} in terms of minority opinion spreading. The subject of conspiracy formation is becoming of strategic importance at different levels of the globalized society. Of special interest is the shift in public opinion that the presence of conspiracy theories can activate \cite{bib3}. Therefore, we search for a robust and generic technique to distinguish between opinions for and against such theories. There are a number of polls and barometers which are questioning the public opinion about conspiracy theories, others about the usage of violence in political and social conflicts, and there are many other indicators of political attitudes. However, it is difficult to find a way to combine these. 
We propose a novel approach framed in terms of information theory and entropy to tackle the issue of the propagation of conspiracy theories. We start with the report of an terroristic event (such as the 9/11 terroristic attack in the USA) represented as a series of individual strings of information denoted respectively by a two-state variable $E_{i}=\pm1, i=1,…, N$.  An official report would be characterized by value $E_{i}=-1$ assigned to all strings, resulting in the order parameter $E=\overline{E_{i}}=-1$, i.e. a fully ordered state, with minimum entropy. This would result in a clear and unambiguous reading and understanding of the event. However, conspiracy theorists might seize this initial presentation based on the series of information strings $E_{i}=-1$ to question their content and changing their meaning (i.e. from $-1$ to $+1$). 
A number of articles would then appear propagating different level of conspiratorial beliefs and accompanied with the comments of the readers who argue in favor or against the new content. The entropy of such textual corpuses is of non-minimum value. Beyond some threshold value, which by simplicity we choose equal to $E=0$, meaning $N/2$ variables with $E_{i}=1$, i.e. probability $p_{i}=1/2$, a total doubt is prevailing in the reading of the event. On this basis, knowing the evolution of the associated entropy offers a way to measure the degree of penetration of a conspiracy theory. 
Our general framework relies on online content made voluntarily available by crowds of people, in response to some news or blog articles published by official news agencies. We apply different aggregation levels (comment, person, discussion thread) when analyzing the change of entropy of the entire corpus.
As opposed to studies where the interactions between agents are more or less complex versions of contagion, the opinion dynamics uncovered by our study displays a variety of interactions that differ by the number of participants, their attitudes and the nature of their arguments. In particular some of the arguments have a more information-like logical character while others originate in, or address, subjective states and experiences. Thus the modeling of the debate process and its outcome will have to transcend the mechanical models of percolation time and yet include elements precise enough to allow quantitative evaluations and predictions.
In particular the information theoretical aspects are only a part of the debate dynamics, aspects which have to do more with the message’s language, positioning and connotations turn out to be equally relevant. The data used for this pilot study are introduced in Sections \ref{sec:2} and \ref{sec:3}. The results have to be taken as qualitative information used for the construction of the model, and not as its quantitative confirmation. Once enough data is gathered, and the proposed model is developed to the level of an application, we hope to obtain more quantitative results. We are thus proposing the first steps for a formalism that parallels the Shannon entropy for communication in order to encode the amount of congeniality of the commentaries with conspiracy theories.

\section{Background information about 9/11 conspiracy theories }
\label{sec:2}

\begin{figure}
  \includegraphics[scale=0.8]{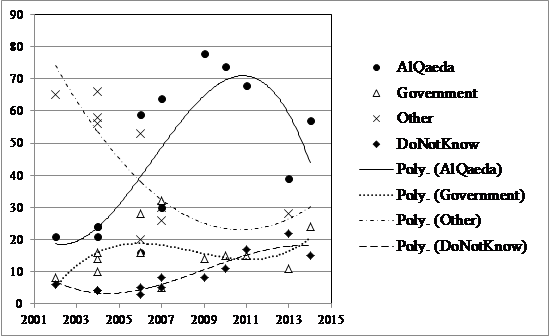}
\centering
\caption{The illustrative fitting lines (2nd and 3rd order polynomial) show dynamics of opinion polls concerning the question “Who did the 9/11 terrorist attacks?” in the US. In the first years after the attack, the majority was not convinced that Al-Qaeda was responsible for the attack. This however has changed and the confidence peaked in 2009; since then it started dropping while at the same time more and more people declare that they do not know who did it.
}
\label{fig:Fig1}      
\end{figure}

After a terroristic attack, a society has to interpret it. Among other goals of violent attacks in the context of political conflict, one is to affect a broader audience beyond the physically targeted victims. People use various cognitive resources to explain an unexpected and disturbing event like this. They are searching for information and arguments that can explain the event, while, at the same time, they seek for the explanation that will help them to protect themselves as much as possible from further attacks. This might be interpreted as teleological (consequentialist) ethics, which derives duty or moral obligation from what is good or desirable as an end to be achieved. It is opposed to deontological ethics, which holds that the basic standards for an action’s being morally right are independent of the good or evil generated. 
Such conditions open a path for the diffusion of conspiracy theories. Just recently, after the terrorist attack on the French satirical weekly magazine Charlie Hebdo and the Kosher supermarket in Paris in January 2015, with 17 people being killed, the results of the public opinion poll \cite{bib4} have shown that in the weeks after the attack a significant number of French citizens held conspiratorial beliefs about it $(17 \%)$. 
Following the instructions and advice by Osama Bin Laden \cite{bib5} in 2002/3, presently active terrorist groups realized that the targeted attacks have a better chance to win support of possible followers, while violence against innocents usually hurts public support. Similarly, an inadequate official response could guide public opinion to their advantage. Conspiracy theories on their own can have an additional leveraging effect: in case they are successful, and the violence cannot be clearly attached to the terrorist, the entire population of peaceful political activists may decide to give an advantage to the terrorist supporters. 
The conspiracy theories that we analyze in this paper concern the September 11 attack, when al-Qaeda terrorists killed 2,996 people and caused at least $ \$ 10$ billion in property and infrastructure damage to the United States. It is rather well known that there was a vast number of conspiracy theories which developed after this attack \cite{bib6}. 
There is no consistent polling about the popularity of this theory. However, we will analyze the available data from official opinion polls, but we will also use opinion mining tools and extract data from the internet about the latest events. An example of the dynamics of the polled opinion about 9/11 conspiracy worldwide is given in Figure \ref{fig:Fig1} \cite{bib7}.

We should mention that the first advertiser of a 9/11 conspiracy theory in France was Thierry Meyssan in his book \cite{bib8}. The book was criticized by the centre and left oriented publishers, such as Libération, Le Nouvel Observateur, etc. However, the rumors hold similarly as in the US. Unfortunately, in France, there are much fewer official polls about this conspiracy \cite{bib9}, \cite{bib10}, \cite{bib11}, \cite{bib12}. The WorldPublicOpinion.org poll \cite{bib10}, conducted during the summer of 2008, revealed that about $63\%$ of the polled French citizens think that for "Al Qaeda" was behind the 9/11 attack,$ 8\%$ think it was "the US government", $7\%$ said "Other" and $23\%$ said they “Do Not Know”. However, the same poll, in which 17 nations were polled, found that majorities believe al-Qaeda carried out the attacks in only 9 of the 17 countries.
We, unfortunately, were not able to collect sufficient number of polls to create valid statistics about the public opinion either in France or in the rest of the world. However, even from the limited data, we can conclude that the spreading of this conspiracy was a slow process which took a couple of years before it became more or less stable. It requires frequent and prolonged monitoring before it could be fully understood.
Therefore we suggest an alternative way to search for opinions about this particular conspiracy.

\section{Analysis of the web-based comments about possible 9/11 conspiracy }
\label{sec:3}

The internet is a very convenient place for conspiracy theories. It provides many so called “proofs” that conspiracy theories rely on, and which are used to convince others of their validity. Therefore we opt to think that the conspiracies and their followers are more likely to express themselves through on-line tools then with conventional opinion polls.

\subsection{Example No. 1, BBC news article}
\label{sec:31}

We analyze comments of the readers of the BBC site to an article that BBC published on  conspiracy theories that refers to some of the conspiracies about September 11 \cite{bib13}. There were 755 comments of the readers. The content of the article and the content of all the comments are presented by the two semantic clouds in Figure \ref{fig:2a} and \ref{fig:2b} respectively. The semantic clouds were constructed using the software Iramuteq \cite{bib14}. The program selects active words (i.e. nouns, verbs, adjectives) and creates a graph based on the words co-occurrences. Comparing the two clouds, one can see that the readers frequently commented on two subjects: “collapse” (of the building next to the Twin towers that were hit by the terrorists) and “person” (who might believe in a conspiracy theory). These two clusters of words create two sides of a conspiracy theory. The “person” cluster implies that the political interests and our own ‘human nature’ are some of the possible incentives for the conspiracy theory. The second cluster of words reflects on the conspiracy theories listed in the BBC article and the discussions about the 9/11 ‘facts’ and whether they match with the reality.

\begin{figure}
\centering
\begin{subfigure}[b]{0.5\textwidth}
                \includegraphics[scale=0.3]{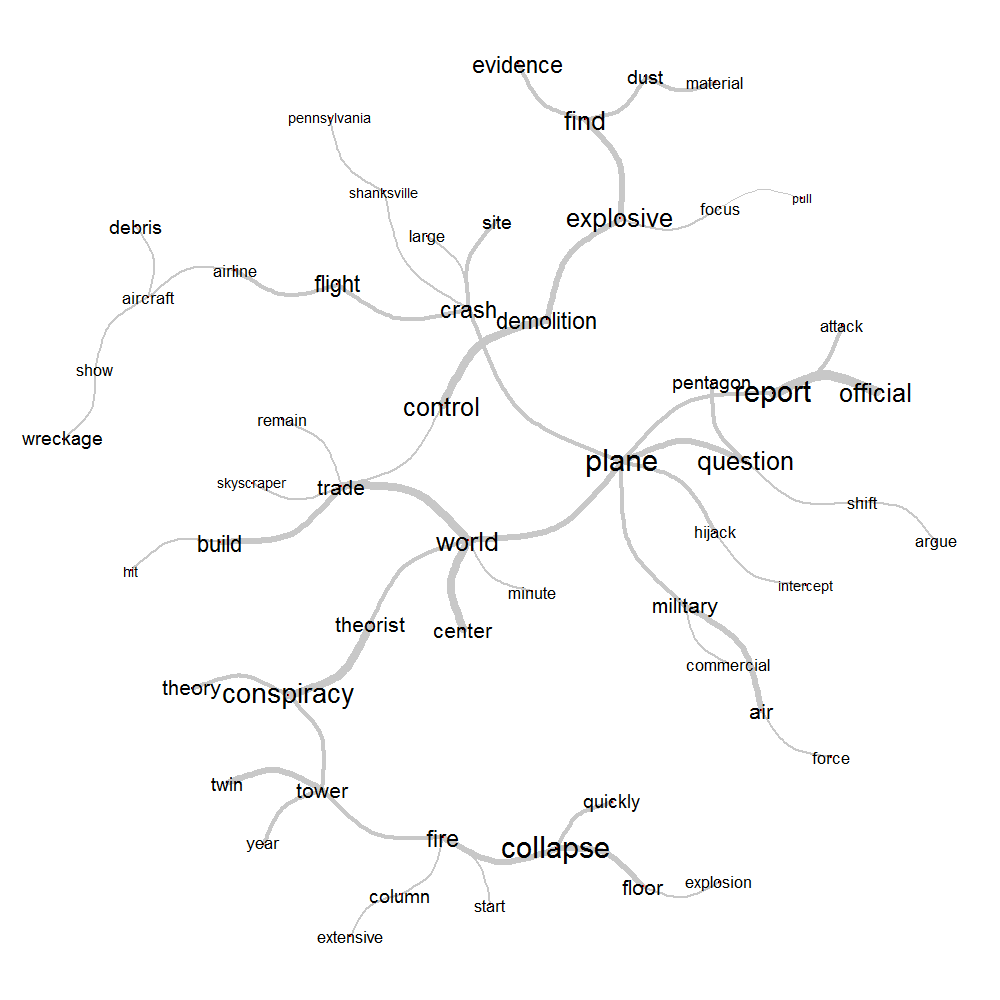}
	\caption{The main text.}
                \label{fig:Fig2a}
        \end{subfigure}%
\begin{subfigure}[b]{0.5\textwidth}
                \includegraphics[scale=0.3]{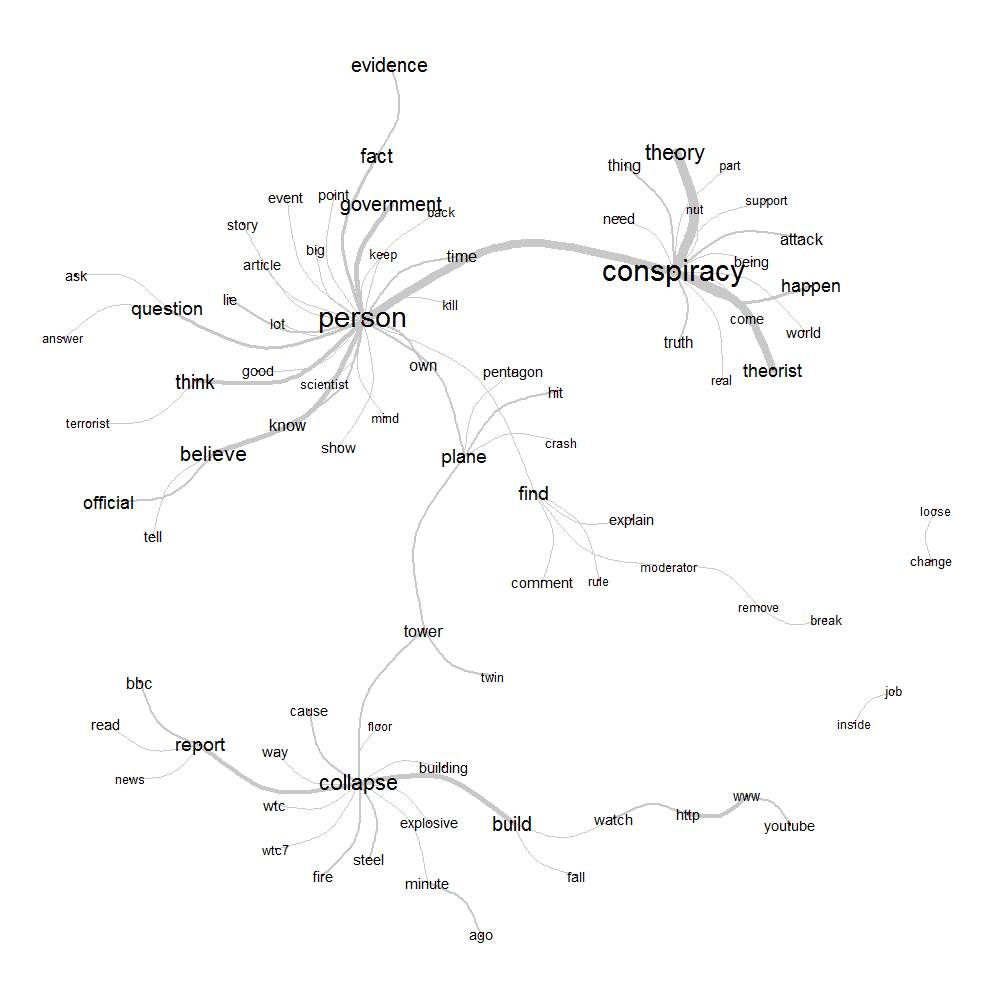}
	\caption{The comments.}
                \label{fig:Fig2b}
        \end{subfigure}%
\caption{(a) Semantic cloud showing the subjects that are discussed in the BBC article: conspiracy, plane, collapse etc. (b) Semantic cloud showing the subjects that are discussed in the comments on the BBC article: conspiracy, person, collapse etc. The edges represent a selection the co-occurrences of those words between comments (frequency > 10).
}
\label{fig:Fig2}      
\end{figure}

\begin{figure}[t]
  \includegraphics[scale=0.6]{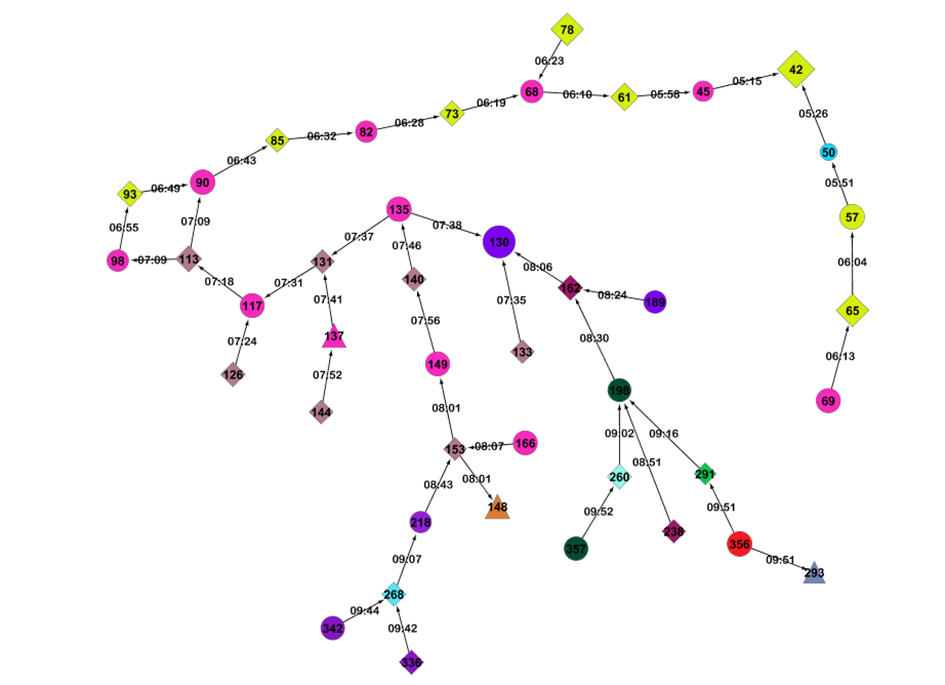}
\centering
\caption{Thread of comments that mention one to the other (establishing the possibility that they share a common ideology) from the BBC article. Comments in the same color were posted under the same pseudo name. The size of the nodes reflects the rating (number of “likes”) by the readers (the larger node, the better the rating). The numbers on the edges give the time when a comment was posted. The shape gives their position: a circle for the ones supporting conspiracy, diamonds for the ones opposing it and a triangle for some comments that are vague or undecided. 
}
\label{fig:Fig3}      
\end{figure}

We further attempt to establish a connection between the comments that refer one to the other. The longest thread of comments in our sample is shown in Figure \ref{fig:Fig3}. We look further at the interactions in this thread of messages and we would like to know whether these messages are supporting or opposing a conspiracy theory. We need this in order to determine the possible interaction patterns of a person who propagates a conspiracy theory.

The selected network of comments shown in Figure \ref{fig:Fig3} starts with the comment No. 42: {\em “In the 50s, the CIA did a lot more questionable things and had more power than they do now, and they got 'outed' for much of it. Nixon couldn't even keep a break-in to an office a secret, I'm supposed to believe a massive coverup with thousands of conspirators and nobody's said *anything*? Soldiers were leaking secrets on Lady Gaga CD's! The government isn't made up of mind-controlled zombies.” }
The community defending conspiracy theories is giving some reasonable and highly ranked/liked arguments. Such is, for example, comment No. 130: {\em “I have no agenda other than to find the truth. Not all CTs as you are calling them are lunatics. Many people wouldn't call themselves CTs but are still not sure we have heard the whole truth. Building 7’s collapse is very strange and unique and there do seem to be many unanswered questions throughout. Stop attacking people asking questions. Being closed minded is not a virtue.”}.

Another selected network of comments (can be found in the Appendix), is the star-shaped one, starting with the comment No. 436: “The one thing that made me sit up and wonder, were the pictures of the Pentagon before it collapsed. The hole (allegedly made by the plane) was just not big enough, and after all these years I still haven't seen a good enough explanation for that anomaly.” 

These examples bring out the two possible main motifs to believe in the conspiracy theory. The first theorist points out the political incentives and the ‘human nature’ of the conspiracy. The second is more concerned about the ‘facts’ and whether they match with the expectations.

\begin{figure}
  \includegraphics[scale=0.7]{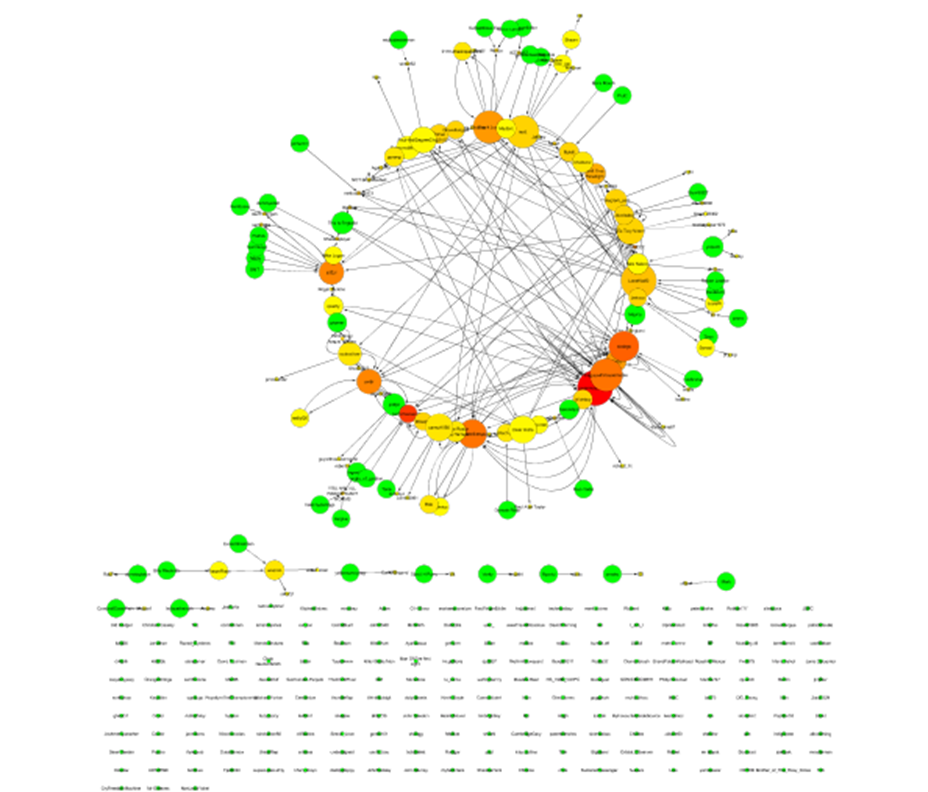}
\centering
\caption{Network of persons who commented on the BBC article on 9/11 conspiracy. The size of the nodes marks the Out-degree, while the colors correspond to In-degree. It shows that the In- and the Out-degree are fairly symmetric. The nodes that are largest (Out-degree/the number of posted comments) are characterized by the strong red color (highest In-degree, the number of comments that a person received).
}
\label{fig:Fig4}      
\end{figure}

In order to understand in how far the commentaries section reflects the opinion of an average reader of the BBC site, we need to know some more information about the activity patterns of the commentators. For this reason, we analyze the emergent social network of the persons who commented on the given BBC article, which is shown in Figure \ref{fig:Fig4}. We distinguished 341 unique pseudo names that appear in the discussions. The network structure in Figure \ref{fig:Fig4} shows that a small-world network is created instantly (as all the comments were posted during a single day). This is also confirmed by the measured degree distribution, which very accurately follows a power law with the slope of $-1.3$.

\begin{figure}
\centering
\begin{subfigure}[b]{0.33\textwidth}
                \includegraphics[scale=0.5]{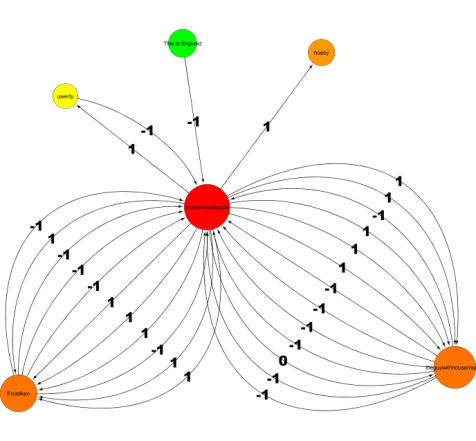}
	\caption{}
                \label{fig:Fig5a}
        \end{subfigure}%
\begin{subfigure}[b]{0.33\textwidth}
                \includegraphics[scale=0.5]{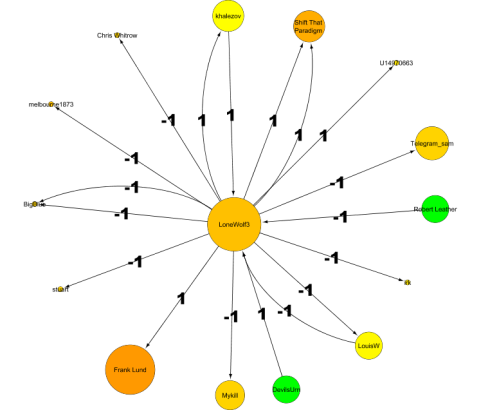}
	\caption{}
                \label{fig:Fig5b}
        \end{subfigure}%
        \begin{subfigure}[b]{0.33\textwidth}
                \includegraphics[scale=0.5]{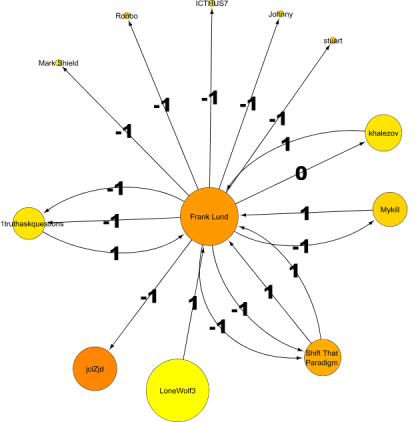}
	\caption{}
                \label{fig:Fig5c}
        \end{subfigure}%
\caption{Sub-networks of the $3$ most active commentators. (a) The first on the left has the largest degree. Most of its comments are supporting conspiracy theories (value 1); the comments that he/she receives are opposing them.  (b) The middle panel shows the second most engaged person; the comments he/she is posting are equally labeled $-1$ or 1, meaning that he/she has not strong opinion about conspiracy theories. This person comments often on the people otherwise isolated.  (c) The third person is against conspiracy theories, commenting always against it (value 1). 
}
\label{fig:Fig5}      
\end{figure}

Figure \ref{fig:Fig5} shows the details of the activity of the three persons who discussed the most frequently. Their patterns of activity are different: the person in Figure \ref{fig:Fig5a}  was mostly engaged with two other persons, while the commentators in Figures \ref{fig:Fig5b} and \ref{fig:Fig5c} had interaction with a large number of persons.

Participants in the discussion are also allowed to give a positive or negative vote to each comment. This results in the final ‘rating’ which is available for each comment on the web site. Unfortunately, we do not have access to the exact procedure used for the generation of rating. An analytical approach of dealing with this problem in the context of crowdsourcing has been discussed in \cite{bib15}. They analyzed a crowdsourcing system consisting of a set of users and a set of binary choice questions under assumption that each user has an unknown, fixed, reliability that determines the user’s error rate in answering questions. However, we have more interest in the analysis of rating per discussion thread then per an individual comment. For this reason, we have analyzed the total rating (sum of ratings of all comments in a discussion thread). This analysis shows that the comments with the highest $(+34)$ and the lowest rating $(-32)$ did not result in long discussions. As a matter of fact, they have remained as individual comments. The statistics also shows that as the thread becomes longer, the probability is higher that the total rating will be positive. By pure chance, it happened so that the longest discussion thread (which is shown in Figure \ref{fig:Fig4}) has also the largest total rating $(+101)$. As we can see in Figure \ref{fig:Fig4}, where the rating is visualized as the size of the nodes, there are 3 comments (No. 42, 78, 130) with large positive rating. These comments might be against or for conspiracy theories as the shape of the comments in Figure \ref{fig:Fig4} implies. As we know that the number of participants in the discussion is $12$, we can conclude that the discussion was interesting also to many other participants and is not a product of mutual support of the discussion participants.

\begin{figure}
  \includegraphics[scale=0.6]{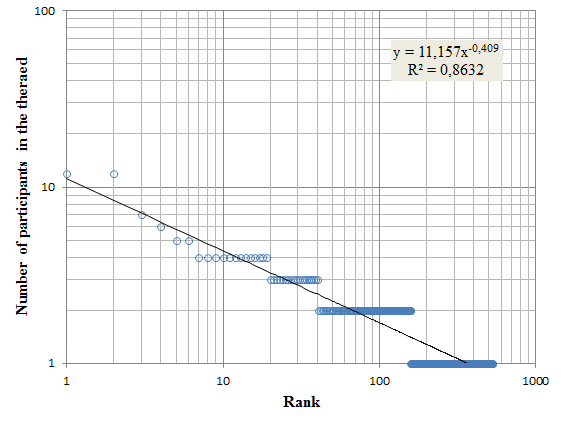}
\centering
\caption{Rank of the size of groups in which people held discussions. It follows a Zipf law of $-0.4$, which is equivalent to a power law of $-2.25$.
}
\label{fig:Fig6}      
\end{figure}

Finally, and most importantly, we analyze the dynamics and the alternation of opinions in the discussion threads: how the discussion impacts the opinion of the people regarding conspiracy theories around 9/11? It is obvious that the discussion usually means a repeated interchange of two different opinions between two or more persons. The most important empirical take away is twofold:
\begin{itemize}
\item The dynamics of the interchange is very fast. All discussions are happening within 1 day with very seldom post-reactions after that. The longest discussion thread is finished within 3 hours.
\item The size of the groups in which people take part in discussions is very small. It is distributed by power law, as shown in Figure \ref{fig:Fig6}. Therefore it is reasonable to assume that all participants in the online discussions have an equal chance to present their opinion to the other participants in that discussion.
\end{itemize}

As opposed to studies where the interactions between agents are more or less complex versions of contagion, the opinion dynamics uncovered by our study displays a variety of interactions that differ by the number of participants, their attitudes and the nature of their arguments. In particular some of the arguments have a more information-like logical character while others originate in or address to subjective states and experiences. Thus the modeling of the discussion process and its outcome will have to transcend the mechanical models of the percolation time and yet include precise enough elements as to allow quantitative evaluations and predictions.

\subsection{Example No. 2, The Telegraph blog}
\label{sec:32}

In our second example, we examine the article written by a historian/writer Tim Stanley and published on The Telegraph Blogs in September 2011. The author justifies the existence and the reasoning of conspiracy theorists by claiming that “conspiracy thinking is a natural part of political discourse”. There are 858 comments in the commentary section. 

\begin{figure}[t]
\centering
\begin{subfigure}[b]{0.5\textwidth}
                \includegraphics[scale=0.3]{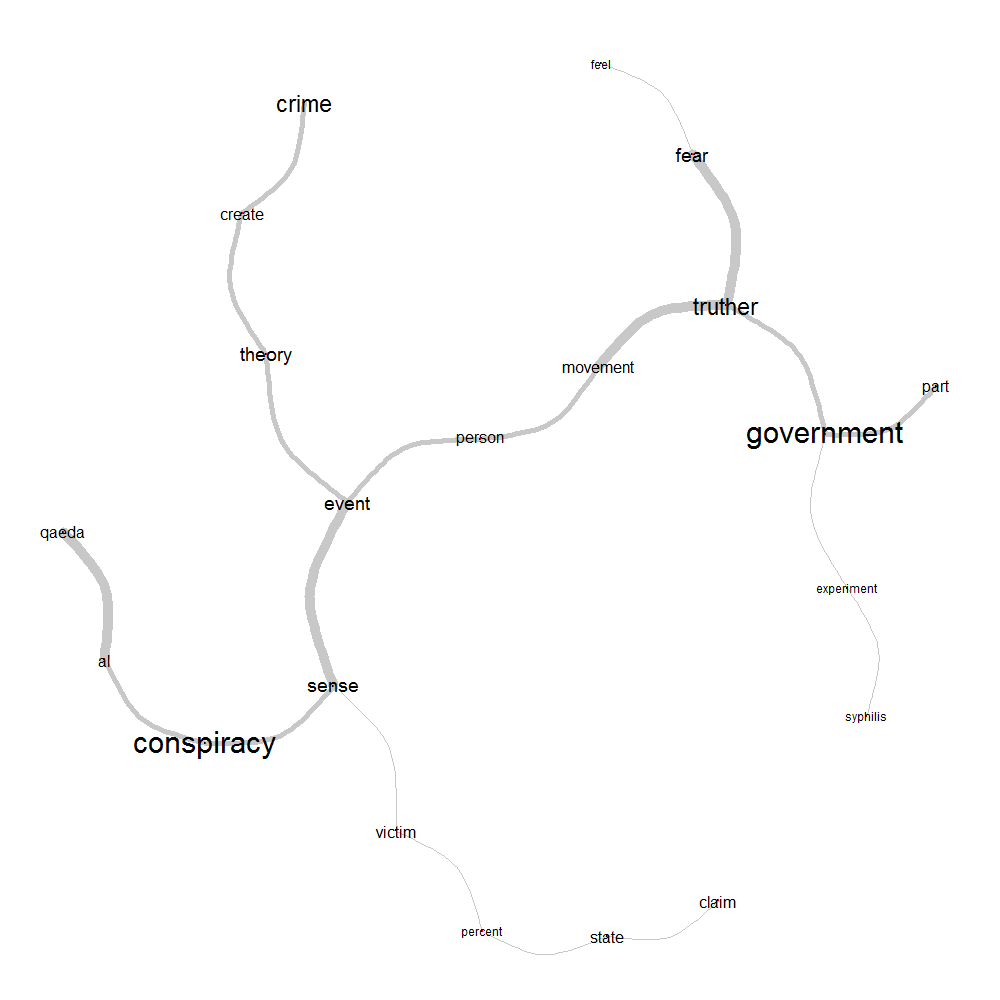}
	\caption{The main text.}
                \label{fig:Fig7a}
        \end{subfigure}%
\begin{subfigure}[b]{0.5\textwidth}
                \includegraphics[scale=0.3]{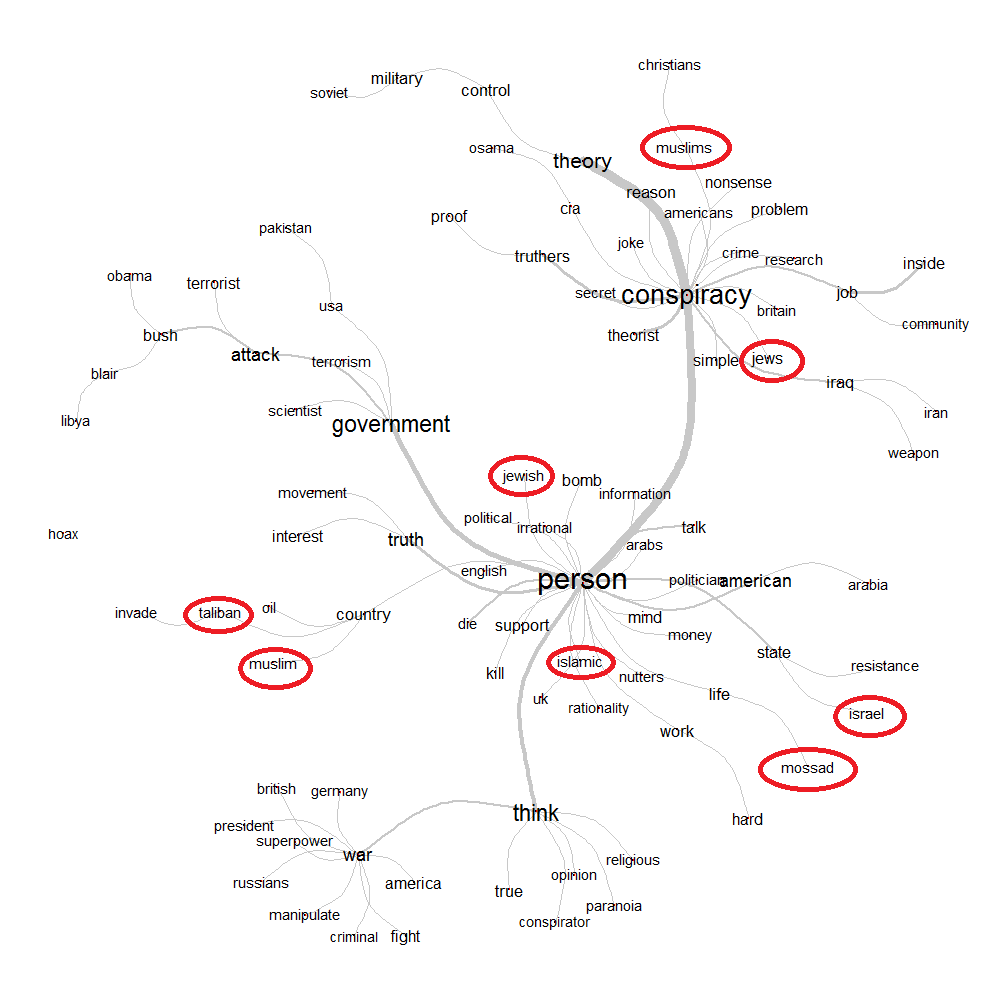}
	\caption{The comments.}
                \label{fig:Fig7b}
        \end{subfigure}%
\caption{(a) Semantic cloud showing the subjects that are discussed in the Telegraph blog. (b) Semantic cloud showing the subjects that are discussed in the comments on the Telegraph blog: conspiracy, fact, person etc. The edges represent the words co-occurrences. The cloud demonstrates the symmetry of arguments used in the comments:  Muslims, Christians and Jews and other similar words branch out from the main nodes in a very symmetrical way. 
}
\label{fig:Fig7}      
\end{figure}

The content of the comments is represented by the semantic cloud in Figure \ref{fig:Fig7}. Obviously, the main tags are similar to the ones in the first example in Figure \ref{fig:Fig3} (conspiracy, person). We need to underline that although the subject of the Telegraph blog is similar with that of the BBC article, the tone is different. It is more ‘philosophical’ in looking for the human side of the conspiracy and less concerned about the facts about it. As such, it is more general and allows a larger variety of opinions to be discussed. 

\begin{figure}[t]
  \includegraphics[scale=0.6]{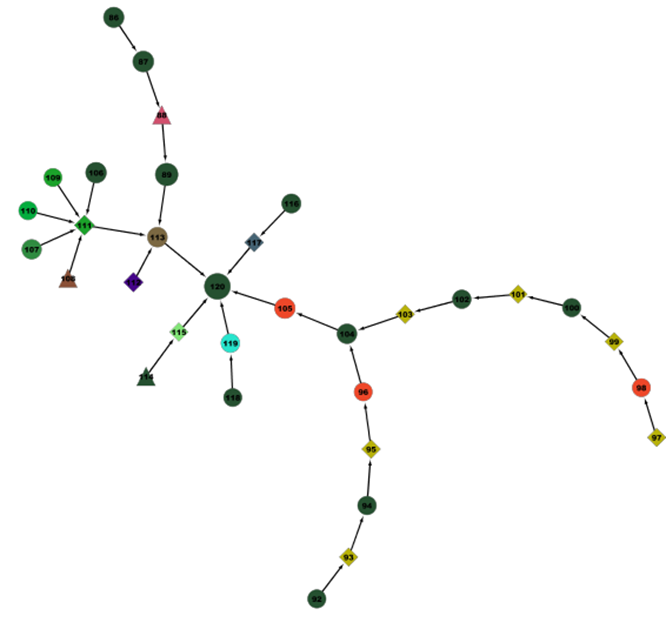}
\centering
\caption{One thread of comments from the Telegraph blog. The colors, size and shape of the nodes have the same meaning as in the BBC example (Figure \ref{fig:Fig3}).
}
\label{fig:Fig8}      
\end{figure}

The comments on the blog site have somewhat different structure then the comments in the previous example (news site). For example: even though the number of comments is larger, the number of persons involved is smaller. In this thread of comments we identified 222 different pseudo names.
The bloggers are more familiar one with another then the discussants of the BBC article, and their readers are perhaps also more used to take part in discussions. This is obvious from, for example, the comment 145: {\em “I've been rude about plenty of Tim's [author of the blog] blogs so I ought to say that this one is really quite illuminating. One thing I'd like to know though is whether Truthers favour one side of the political spectrum, or both, or none.”}
Therefore, the discussions take somewhat different flow then what we saw in the Example $No. 1$. The discussion threads are longer than the ones that appeared in response to the BBC article, and the isolated comments (without any reaction) are less frequent. One example of a thread of comments is given in Figure \ref{fig:Fig8}.

\begin{figure}
  \includegraphics[scale=0.5]{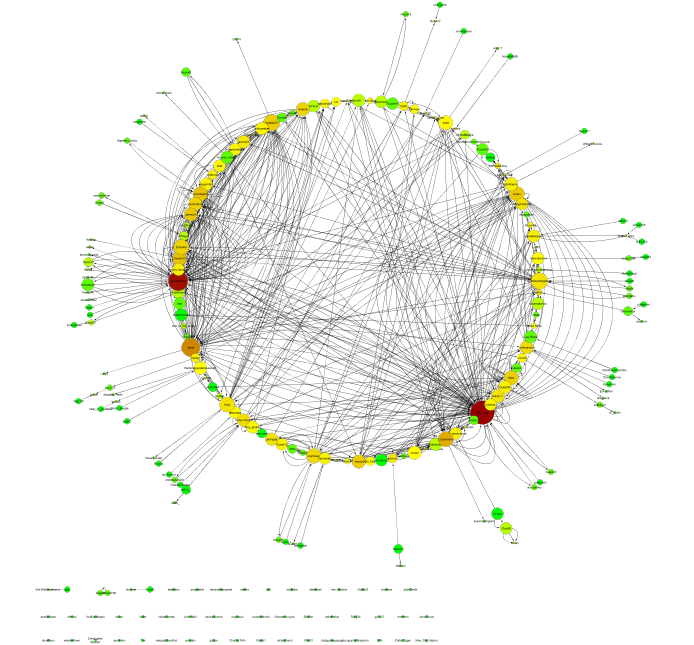}
\centering
\caption{Network of discussants who commented on the Telegraph blog on 9/11 conspiracy. The size of the nodes marks the Out-degree and the colors correspond to In-degree. It demonstrates that the In- and the Out-degree are correlated: the red nodes, which have the largest In-degree are at the same time the ones that have the largest Out-degree (as the nodes are the largest in size).
}
\label{fig:Fig9}      
\end{figure}

\begin{figure}
\centering
\begin{subfigure}[b]{0.33\textwidth}
                \includegraphics[scale=0.7]{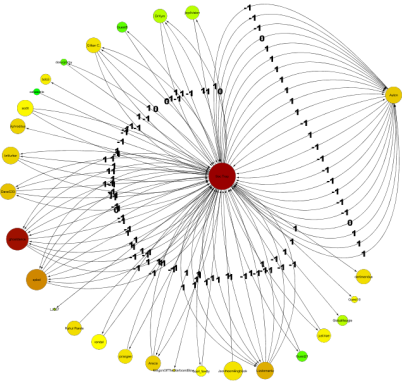}
	\caption{}
                \label{fig:Fig10a}
        \end{subfigure}%
\begin{subfigure}[b]{0.33\textwidth}
                \includegraphics[scale=0.7]{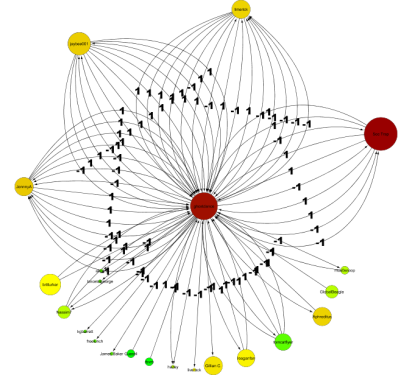}
	\caption{}
                \label{fig:Fig10b}
        \end{subfigure}%
        \begin{subfigure}[b]{0.33\textwidth}
                \includegraphics[scale=0.7]{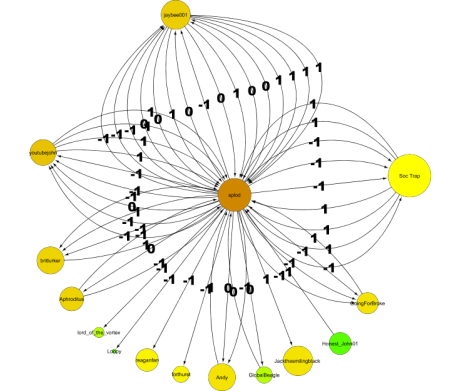}
	\caption{}
                \label{fig:Fig10c}
        \end{subfigure}%
\caption{Sub-networks of the most frequently engaged commentators in The Telegraph blog. (a) The first on the left has the largest degree. Most of its comments are supporting conspiracy theories; the comments that he/she receives are opposing them. (b) The middle panel shows the second most active person; the comments he/she posting are mostly labeled -1, meaning that he/she is opposing conspiracy theories. (c) The third person, same as the second one, opposes the conspiracy theories.
}
\label{fig:Fig10}      
\end{figure}

The network of discussants, Figure \ref{fig:Fig9}, shows typical characteristics of a small-world network. The IN- and OUT-degrees are quite symmetrical and the degree of the nodes follows power law with the slope $-1$. The activity patterns of the most active persons are visualized by their sub-networks in Figure \ref{fig:Fig10}. 

\begin{figure}
  \includegraphics[scale=0.7]{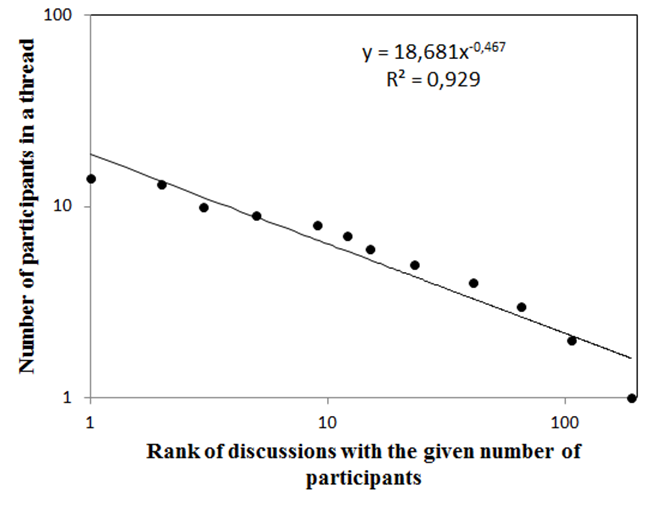}
\centering
\caption{Rank of the size of groups in which people held discussions. It follows a Zipf law of -0.48, which is equivalent to a power law of -2.08.
}
\label{fig:Fig11}      
\end{figure}

Finally, we examine how many people participate in a discussion thread. The result (Figure \ref{fig:Fig11}) shows that even when the length of the thread is very large, the number of participants in each discussion is very small. The slope of the Zipf/Pareto law fitting line $(\sim 0.5)$ tells us that the probability that a discussion will engage a more then $x$ participants scales with $x$ to the power of approximately $-2$ $\big( p[x>X] \sim x^{-2} \big)$.

\subsection{Observations common to both experiments and their relation with existing literature}
\label{sec:33}

A social psychological study of online discussion of 9/11 conspiracy theories \cite{bib16} brought up several observations, defining some characteristics of a person involved in those dicussions. They found out that the conspiracist commenters are more likely to argue against opposing interpretation and less likely to argue in favor of their own interpretation, while the opposite would be true for conventionalist commenters. In addition, conspiracists are more likely to express mistrust and make more positive and fewer negative references to other conspiracy theories. They also indicate that conspiracists are largely unwilling to apply the “conspiracy theory” label to their own beliefs and object when others did so, lending support to the long-held suggestion that conspiracy belief carries a social stigma. Finally, conventionalist (defenders of the official report) tend to have a more hostile tone. These tendencies in persuasive communication can be understood as a reflection of an underlying conspiracist worldview in which the details of individual conspiracy theories are less important than a generalized rejection of official explanations.

We generally agree with the above findings. More specifically, we confirm that the rationalist arguments tend to have a more hostile tone. We also agree that the conspiracist do not change their view even after exposed to very clear arguments, which confirms their generalized and unanimous rejection of official explanations.

When it comes to the analysis of rating (the number of ‘Likes’), we refer to the work of Muchnik et al in \cite{bib17}, who discovered that prior ratings create significant bias in individual rating behavior, and positive and negative social influences created asymmetric effects. In other words, the voting is path dependent. We actually find out that the comments with a negative rating have fewer responses then the comments with a positive rating. The most popular comments (with a high probability to promote conspiracy theories) are also the most frequently contradicted ones. 
Further, our research brings out that the on-line discussions about conspiracy theories are held in small groups of people who are exchanging their opinions. Maximum size of the discussion group is $14$ and usually the discussion groups are much smaller, often containing only a single opinion. The size of the groups follows a power law approximately, with the slope of $-2$. 

The emergent social network of the commentators shows small-world network properties, characterized by a power law distribution of the commentators’ activity. The In- and Out-degrees, i.e. the number of the received and the number of the posted comments per person are fairly balanced. 
The data indicates that the dynamics of the collective opinion is determined not only by the informational content of the debate, but aspects which have to do with the semantics and the network connectivity of the message turn out to be equally relevant. Once enough data will be gathered, we will disentangle those aspects and evaluate further their effects. 

\section{Estimation of entropy of the comments on the articles reporting conspiracy }
\label{sec:4}

Entropy is a measure of the information content, or rather, its uncertainty. It can be assessed in various ways and here we propose a possible approach.
Let $E_{i}$ stand for the state variable, $E_{i}=-1$ (true) or $1$ (false), $i=1,…, N$.  Assigning $E_{i}=-1$ to all strings, the resulting order parameter $E=-1$ marks a fully ordered state, with minimum entropy $(H=0)$. 

Conspiracy theorists comment to the report, focusing on one string $ E_{k}$ and changing its meaning (i.e. from $-1$ to $+1$). Entropy here characterizes our uncertainty about our source of information, and increases with adding more comments of greater randomness. The source is also characterized by the probability distribution of the samples drawn from it. The idea behind the Shannon entropy is that the less likely an event is, the more information it provides when it occurs. The amount of information contained in a string $E_{i}$ is not equally weighted in the formation of the opinion about conspiracy theory defined by the series of $E_{i}$’s. The expected value (average) of the information changes with each comments received. The entropy of the entire series $E_{i}$ is therefore given by:

\begin{equation}
H= - \sum_{i=1}^n p_{i} log_2 (p_{i}),
\label{eq:Shannon}
\end{equation}

where $p_i$ are coefficients that define different significance of the information provided in the strings $E_i$ and belongs to a probability distribution which satisfies the condition:

\begin{equation}
\sum_{i=1}^n p_{i} =1.
\label{eq:sumpi}
\end{equation}

It is far from trivial to bring the above ideas to realization and provide an automatized procedure for the measure of entropy and the estimation of the discrete probabilities $p_{i}$. 

In the subsequent subsections we will apply as the data previously introduced in the BBC and The Telegraph blog examples in an attempt to estimate $p_{i}$. The data might be organized is several ways, as it is illustrated in Table 1 and will be discussed in the sequel.

\begin{center}
\begin{tabular}{| l | l | l | l | l || l |}
\hline
  & Discussion & Discussion & ... & Discussion & $p_{i}$ per\\
    & thread 1 & thread 2  & & thread Z  & person \\
    \hline
Person & (-1, 0, 1) & (-1, 0, 1) & ... & (-1, 0, 1) & $p_{i} \sim \overline{E_{i}}$\\
 1  &  &  &  &  & $=f(INdegree)$\\
\hline
Person & (-1, 0, 1) & (-1, 0, 1) & ... & (-1, 0, 1) & $p_{i} \sim \overline{E_{i}}$\\
 2  &  &  &  &  & $=f(INdegree)$\\
\hline
... & ... & ... &  & ... &...\\
\hline
Person & (-1, 0, 1) & (-1, 0, 1) & ... & (-1, 0, 1) & $p_{i} \sim \overline{E_{i}}$\\
 W  &  &  &  &  & $=f(INdegree)$\\
\hline
\hline
$p_{i}$ per & $p_{i} \sim \overline{E_{i}}=$ & $p_{i} \sim \overline{E_{i}}=$ & ... & $p_{i} \sim \overline{E_{i}}=$ & \\
thread         & $f(Thread$ $Size)$                               & $f(Thread$ $Size)$   &                                      & $f(Thread$ $Size)$                                 & $H \sim - \sum p_{i} log(p_{i})$  \\ \hline
\end{tabular}
\end{center}
Table 1: Observed content of comments could be organized according to persons or discussions. Both rows and columns are expected to satisfy a power law  distribution, which is a statistical simplification that in the next stages of this work might be used exactly the other way around, i.e.  in order to predict/confirm whether the mesages are supporting or opposing conspiracy theory.

A common way to define entropy for a text is based on the Markov model which indeed might also appropriate for the comments-level entropy. This approach will be considered in sub-section \ref{sec:41} when will monitor the average value of $E_i$ of all previously (and randomly) posted comments. We however are looking for a model that would include the aspects which have to do with the properties of networks of messages and persons. Therefore we vary the approach to the “discretization” of the system or rather, to the aggregation of individual comments. In sub-section \ref{sec:42} we estimate the average value of $E_i$ per discussion thread. In the third case, sub-section \ref{sec:43}, we monitor the average $\overline{E_{i}}$ value per person. 

Combining these distributions of the $\overline{E_{i}}$ measured as a function of discussion thread or a person IN/OUT-degree with the previously estimated distributions of the discussion size and the persons connectivity, it would be possible to estimate $p_{i}$.

An interesting model that applies minimax entropy model to improve the quality of the noisy labels obtained during crowdsourcing acquisition has been recently reported in \cite{bib18}. They have assumed a probability distribution over workers, items and labels. The minimax entropy approach is assuming the ground truth from the minimum entropy and generation the entire probability distribution of the worker-item label models from the maximum entropy. To make a remote parallel with such model, we could compare workers with discussants, items with the comments under discussions and labels with rating. We intend to work on this formalization in the future. In the current paper we develop the concept and we empirically validate many aspects of the procedure. However, a full integration of those basic steps in a coherent and feasible procedure has to be developed yet.

\subsection{Monitoring the state variable of comments as they are appearing (in time) }
\label{sec:41}

Before we start with the analysis of the comments, the order variable value $\overline{E_{i}}$ of the original article/blog needs to be estimated. The BBC article is listing different 9/11 conspiracy theories in a factual style. This article is critical about conspiracy theorists, as the following sentence copied from this article implies: {\em “Numerous official reports have been published since the Twin Towers fell, but just when a piece of evidence casts doubt on one theory, the focus then shifts to the next "unanswered question"”}. Therefore the value of its order variable $\overline{E_{i}}$ is negative. The Telegraph article is also listing a number of conspiracy theories, but in a less judgmental style {\em (“…in fact conspiracy thinking is a natural part of political discourse. It represents an effort to make sense of apparently senseless events. People conquer their fears by drawing connections between unconnected tragedies to create a unified theory that brings order out of chaos.”}). Therefore this article has a positive $\overline{E_{i}}$. Starting from these two different initial conditions, we monitor the dynamics in $\overline{E_{i}}$ of the commentaries.

\begin{figure}
\centering
\begin{subfigure}[b]{0.5\textwidth}
                \includegraphics[scale=0.35]{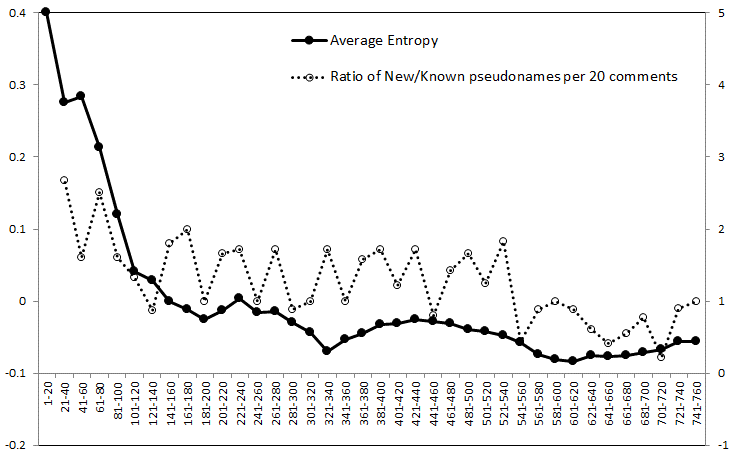}
	\caption{The BBB example.}
                \label{fig:Fig12a}
        \end{subfigure}%
\begin{subfigure}[b]{0.5\textwidth}
                \includegraphics[scale=0.35]{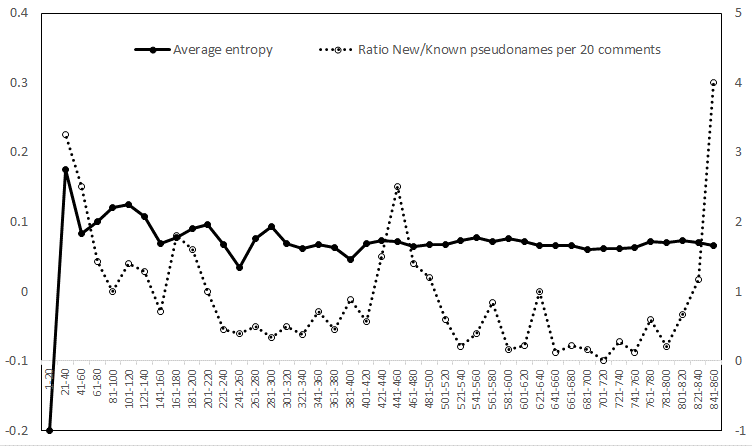}
	\caption{The Telegraph example.}
                \label{fig:Fig12b}
        \end{subfigure}%
\caption{(a) The figure shows the $\overline{E_{i}}$ dynamics over the entire corpus of 775 comments on BBC article. Each black circle (full line, primary axis on the left) represents an incremental average value recalculated after the constant size window of 20 comments. The white circles (dashed line, secondary axis on the right) represent the ratio of the new over the previously involved commentators, for each set of 20 comments. (b) The figure shows the dynamics of $\overline{E_{i}}$ over the entire set of 858 comments on The Telegraph article. Each black circle (full line, primary axis on the left side) represents the average incremental value recalculated after the constant size window of 20 comments. The white circles (dashed line, secondary axis on the right side) represent the ratio of the new and previously involved commentators, for each set of 20 comments.
}
\label{fig:Fig12}      
\end{figure}

Assigning to each comment a value $-1$ or $+1$ (personal judgement of the author) and linking it to one of the original $E_{i}$ strings, to which the comment refers to, the value of the argument posted in the original article deteriorates. After a certain number of comments, the entropy of the system can be recalculated. This process can be continued by taking a measure of entropy after each X comments. To give an impression of the expected dynamics, we apply this procedure with a time step randomly chosen to 20 $(X=20)$. This is shown in Figure \ref{fig:Fig12a} for the BBC example and Figure \ref{fig:Fig12b} for The Telegraph example.

Although those two examples show different behavior, they do show some similarities:
\begin{itemize}
\item	The first set of 20 commentaries is opposing/challenging the opinion promoted in the original article. In the BBC example, whose $\overline{E_{i}}<0$, the first set of comments has an $\overline{E_{i}}=0.4$. In the Telegraph example, average(Ei)>0, the first set of comments has an$\overline{E_{i}}=-0.2$. 
\item	Increasing the number of comments, the order variable bounces back to the opinion/side that the original article has promoted. 
\item	After a number of fluctuations the $\overline{E_{i}}$ saturates to a value of -0.07 for the BBC article and +0.07 for the Telegraph article. 
\item	The process of saturation of the order variable $\overline{E_{i}}$ is accompanied with the low level of interest of the readers (when the ratio New/Known pseudo names is less than 1). 
\end{itemize}

The oscillatory dynamics of $\overline{E_{i}}$ as a function of time is significantly different in the two examples, and therefore we cannot approximate it by a mathematical function necessary for the estimation of $p_{i}$.  We might be able to find the indicators for the ‘jumps’ in the order parameter from the semantic similarity analysis applied to the level of the sets of comments. The string similarity measures such as Jaccard or Tversky distance as well as the Kullback-Leibler distance have already been tested for this purpose and their applicability for this purpose seems to be feasible.

\subsection{Monitoring the state variable of discussion threads }
\label{sec:42}

\begin{figure}
\centering
\begin{subfigure}[b]{0.5\textwidth}
                \includegraphics[scale=0.35]{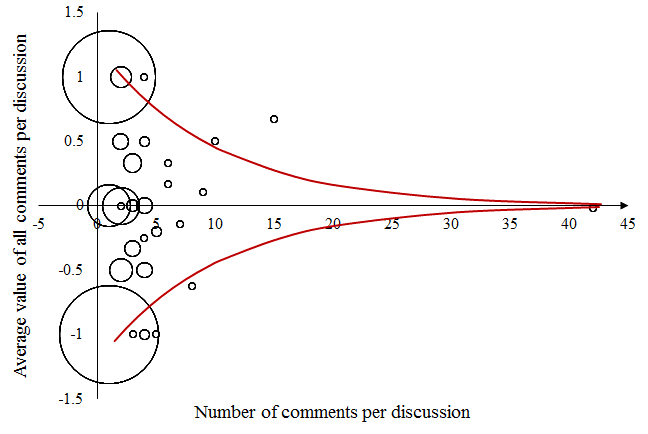}
	\caption{The BBC example.}
                \label{fig:Fig13a}
        \end{subfigure}%
\begin{subfigure}[b]{0.5\textwidth}
                \includegraphics[scale=0.35]{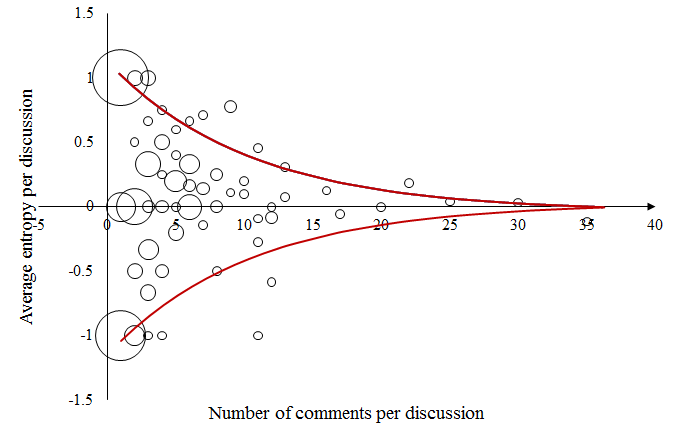}
	\caption{The Telegraph example.}
                \label{fig:Fig13b}
        \end{subfigure}%
\caption{(a) The figure shows the order parameter $\overline{E_{i}}$ as a function of the discussion thread length Li, measured for the discussions on the BBC site. Each circle of the minimum size represents only one discussion (such as the longest discussion of the length 42, which appears only once). The circle of the largest size represents 198 comments for which there was no reaction (-1, against conspiracy theories). The graph is rather symmetrical in the upper and the lower part. Clearly, there is a tendency to have stronger polarization of $\overline{E_{i}}$) for the shorter discussions, while the longer ones are more `neutral`, the resulting order parameter is closer to zero. This tendency is visualized by the red line defined by Eq. \ref{eq:discussions}. (b) The figure shows the distribution of the $\overline{E_{i}}$ of the discussions on the Telegraph example. The circles of minimum size represent only one discussion (such as the longest discussion of the length 35). The circle of the largest size (1, pro conspiracy theories) represents 41 comments to which there was no reaction. The red line is following the exponential law from Eq. \ref{eq:discussions}.
}
\label{fig:Fig13}      
\end{figure}

The length (size) and the balance of opinions in a discussion is determining the impact that the discussion has on the entropy of the system. Single comment would have much less impact then the comments given in a discussion engaging a number of persons. From the power law distribution of the discussion groups, the weight distribution can easily be recalculated and it will also be a power law function. Also, the rating per discussion can be a parameter used for this purpose. To illustrate the shape of the statistics underlying the discussion thread structure, we use the BBC example data, as shown in Figure \ref{fig:Fig13}. The data are very `promising` showing a tent shaped distribution. Assuming that such distribution exists would indeed shorten the entropy calculations significantly comparing with the situation presented as case 1.

The red line in Figure \ref{fig:Fig13} is an approximate border line and it is an exponential function of the discussion length $L_i$:
\begin{equation}
\overline{E_i}= \pm \exp( {\frac{L_i}{const}}).
\label{eq:discussions}
\end{equation}

The constant is chosen such that $L_{i}=1$ returns $E_{i}$ value of $\pm 1$. This border line represent the maximal (absolute) expected value that the $\overline{E_{i}}$  of a discussion. Knowing this relation and knowing the distribution of the discussion thread legths,  the $p_i$ distribution could be recalculated. Of course it would be necessary to apply some sort of normalization (providing $\sum_i p_{i} =1$).

\subsection{Monitoring the state variable of the persons}
\label{sec:43}

Our intuition suggests that the comments of the persons who are only involved in a single discussion would be weighted less then the comments of the persons who are very active and able to constantly provide strong opinion. From the power law distribution of the discussants activity, the weight distribution can easily be recalculated and it will also be a power law function. In this case rating could also be used to select the persons who are able to receive the support of the audience.

\begin{figure}
\centering
\begin{subfigure}[b]{0.5\textwidth}
                \includegraphics[scale=0.3]{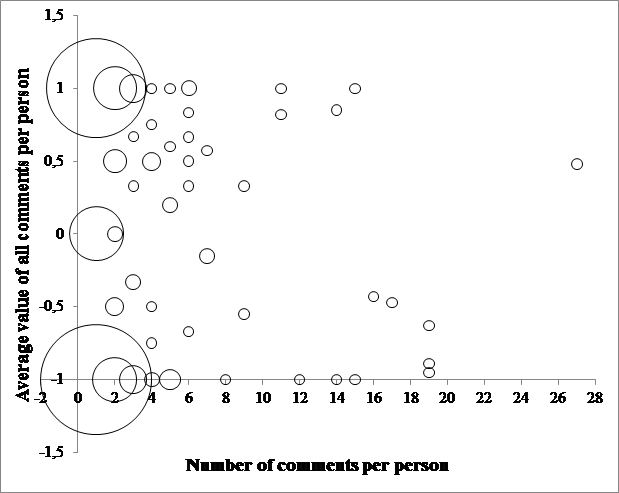}
	\caption{The BBC example.}
                \label{fig:Fig14a}
        \end{subfigure}%
\begin{subfigure}[b]{0.5\textwidth}
                \includegraphics[scale=0.5]{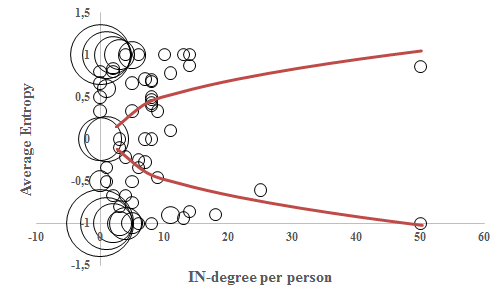}
	\caption{The Telegraph example.}
                \label{fig:Fig14b}
        \end{subfigure}%
\caption{(a) The figure shows the ‘stability’ of discussants` opinion, using the BBC example data. Each circle of minimum size represents the average opinion only one person. The circle of the largest size represents the average opinions of 113 persons (as a matter of fact that persons gave only one comment valued -1 (against conspiracy theories). (b) The figure shows The Telegraphs commentators’ opinion as a function of their IN-degree ratio. 
}
\label{fig:Fig14}      
\end{figure}

We have tested whether it would be useful to add additional weight coefficient to the comments of the persons that are involved in discussions more frequently by measuring the stability of the opinion of the discussants as presented in Figure \ref{fig:Fig14}. The weight factor could be estimated from the data distribution. This graph shows that the discussants are typically very polarized (-1 or 1). This holds for most of the persons. The change of entropy is in such cases almost ‘neutralized’ by the opposite attitudes – as long as the distribution that describes those persons it is symmetric. However, there is also a number of persons whose opinion is not -1 or 1, but somewhere in between. Finally, there is also a number of ‘internet trolls’ who are posting off-topic messages. They are characterized with the $E_i=0$ and the IN-degree=0 and should not be taken into consideration when estimating the change of the order variable.

The red border line in Figure \ref{fig:Fig14} is a power law function of the persons’ IN-degree (number of received comments):
\begin{equation}
\overline{E_i}= \pm \Big( \frac{INdegree}{const} \Big)^{0.5}.
\label{eq:persons}
\end{equation}

The constant in Eq. \ref{eq:persons} equals to the maximum IN-degree of all persons. The red border line in Figure \ref{fig:Fig14} represents the minimum absolute expected value $\overline{E_i}$ of a person.  As a first approximation, and in combination with the distribution of the persons IN-degree estimated in Section \ref{sec:3}, this simplified relation could also be used to determine $p_i$ (after applying some sort of normalization which would provide that $\sum_i p_i=1$).

An interesting side-observation from Figure \ref{fig:Fig14} is that the persons with the largest IN-degree are at the same time the “stubborn” persons who are not changing their original view (and therefore have a highly polarized average $E_i$). It looks as if they get involved in discussions while trying to convince the others to change their opinion. This behavior in the on-line environment is, according to the recent work of Fisher and Keil \cite{bib19}, not surprising. Namely, they claim that the nature and context of argumentative exchange depends on the social context and distinguish two different modes of argumentation: arguing to win and arguing to learn. They experimentally proved that arguing in private prompts a mindset of arguing to learn, while arguing in public prompts a mindset of arguing to win. Further on, they also showed that the choice of person with whom participants chose to argue varied as a function of either a public or private social setting. Namely, there is a preference to argue with the less knowledgeable person in public and with the more knowledgeable person in private. From this perspective, all on-line discussions, being placed in a public setting, although behind the pseudo names, are accommodating ‘argue to win’ discussion modus.

\section{Conclusions}
\label{sec:5}

In this paper, we apply the concept of entropy as a measure of the penetration depth of a conspiracy theory (i.e. a measure for the information uncertainty that the conspiracy theorists introduce through various on-line content). Our basic model and the pioneering empirical studies point out to the necessity to extend the common measures of information with some aspects of the social networks theory and some semantic string similarity measures in order to enable such a qualitative analysis of a large data corpus. In this particular application, we deal with the information which carry subjective and personal rather than purely factual content and contains aspects of mental processing which are not governed by regular Boolean logics. 

More precisely, in addition to the information content measured by Shannon-like methods presented in Section \ref{sec:4} one needs to employ the knowledge of the interaction patterns in order to classify the content of the messages and the traces of conspiratorial beliefs. At the collective level, one should recognize and measure the common vocabulary that labels and singles out /separates certain semantic proximity, symmetry or idiomatic expressions (e.g. revolution vs. violence, jihad vs. terrorism, freedom of expression vs blasphemy, multicultural vs apostate, free thinker / nonbeliever etc). While the use of one word or another does not affect much the information content, it does label and affect the emotional content and dynamics of the interaction. 

Tagging the above signs could be used for machine learning training in order to classify the order parameter of the individual comments. The large training sets produced on the articles of similar content in different social or geographical environments, supported by the underlying assumptions of the predefined statistics based on the social and comments networks can provide information theory measures, and bring entropy concepts in dealing with subjective / social / human dynamics. This is a long overdue move which may largely benefit both the exact sciences as well as the social / human sciences. We hope that our first stage calculations will trigger additional investigations towards a full integration of the various ideas here discussed towards a coherent and feasible procedure.

\section*{Acknowledgments }
We acknowledge all participants of the Workshop “From Opinion Dynamics to Voting, Conflict and Terrorism” held 30/31 March 2015 in Paris for useful interactions. This work has been performed under the DGA Grant «L’impact des supporters passifs dans l’action terroriste : une approche à partir de la sociophysique», coordinator Serge Galam, DGA-2012 60 0013 00470 75 0.

\end{document}